\documentclass{article}

\usepackage{arxiv}
\usepackage{graphicx}
\usepackage[utf8]{inputenc} 
\usepackage[T1]{fontenc}    
\usepackage{hyperref}       
\usepackage{url}            
\usepackage{booktabs}       
\usepackage{amsfonts}       
\usepackage{nicefrac}       
\usepackage{microtype}      
\usepackage{lipsum}

\title{Unconventional Bio-Inspired Model \\for Design of Logic Gates}

\author{
  Theofanis Floros \\
  Department of Electrical and Computer Engineering,\\
   Democritus University of Thrace, \\
  Xanthi, Greece \\
   \And
 Karolos-Alexandros~Tsakalos \\
  Department of Electrical and Computer Engineering,\\
   Democritus University of Thrace, \\
  Xanthi, Greece \\
   \And
   Nikolaos Dourvas\\
  Department of Electrical and Computer Engineering,\\
   Democritus University of Thrace, \\
  Xanthi, Greece \\
     \And
     Michail-Antisthenis Tsompanas\\
     Unconventional Computing Laboratory,\\
     University of the West of England, \\
     Bristol, United Kingdom \\
    \And
    Georgios Ch. Sirakoulis \\
  Department of Electrical and Computer Engineering,\\
   Democritus University of Thrace, \\
  Xanthi, Greece \\   
}

\begin{document}
\maketitle

\begin{abstract}
During the last years, a well studied biological substrate, namely \textit{Physarum polycephalum}, has been proven efficient on finding appropriate and efficient solutions in hard to solve complex mathematical problems. The plasmodium of \textit{P. polycephalum} is a single-cell that serves as a prosperous bio-computational example. Consequently, it has been successfully utilized in the past to solve a variety of path problems in graphs and combinatorial problems. In this work, this interesting behaviour is mimicked by a robust unconventional computational model, drawing inspiration from the notion of Cellular and Learning Automata. Namely, we employ principles of Cellular Automata (CAs) enriched with learning capabilities to develop a robust computational model, able of modelling appropriately the aforementioned biological substrate and, thus, capturing its computational capabilities. CAs are very efficient in modelling biological systems and solving scientific problems, owing to their ability of incarnating essential properties of a system where global behaviour arises as an effect of simple components, interacting locally. The resulting computational tool, after combining CAs with learning capabilities, should be appropriate for modelling the behaviour of living organisms. Thus, the inherent abilities and computational characteristics of the proposed bio-inspired model are stressed towards the experimental verification of \textit{Physarum}'s ability to model Logic Gates, while trying to find minimal paths in properly configured mazes with food sources. The presented simulation results for various Logic Gates are found in good agreement, both qualitatively and quantitatively, with the corresponding experimental results, proving the efficacy of this unconventional bio-inspired model and providing useful insights for its enhanced usage in various computing applications.
\end{abstract}


\section{Introduction}
\subsection{Living Organisms Inspiration}
Several organisms develop a form of interconnected networks as part of their food foraging strategy and discovering new resources for exploitation of the available space. Such organisms constantly adapt to their environment and have to balance the cost of creating 
an effective network with the consequences of possible failure in an ever competitive world. Unlike human infrastructure systems, these biological systems have evolved through continuous and repetitious evolutionary procedures and have reached an equilibrium stage between cost, efficiency and endurance. In the past decades, taking continuous inspiration from biology, computer science has managed to introduce useful problem solving approaches such as neural networks, genetic algorithms and optimization algorithms developed by observing, for instance, ant colonies~\cite{Bitsakidis2017} (ant colony optimization \cite{Blum2005}). In this work, we consider the properties of the unicellular slime mould - \textit{P. polycephalum} - during the creation of both adaptable networks~\cite{Tero2010,adamatzky2012,tsompanas2014,tsompanas2015cellular,Evangelidis3D} and logic gates~\cite{Adamatzky2010}, which has led to the development of a biologically inspired model.

\subsection{Logic Gates implementation with Physarum polycephalum}

The plasmodium of \textit{P. polycephalum} is a single-cell with multiple diploid nuclei, visible to the naked eye and capable of being expanded to a scale of meters. When the plasmodium is placed in an environment with distributed nutrients, it develops a network of protoplasmic tubes with length equal to the distance between the sources of food~\cite{Adamatzky2010}. While foraging and transporting nutrients throughout its protoplasmic body, plasmodium converges to the minimum path in a maze~\cite{Nakagaki2001,Adamatzky2012b}, calculates planar proximity graphs~\cite{Adamatzky2009} as well as plane tessellations~\cite{Shirakawa2009}, discloses an early form of memory~\cite{Smith2017} and performs basic logic operations~\cite{Adamatzky2010}. The plasmodium can be considered as a general purpose computer because it is capable of simulating the Kolmogorov-Uspensky machine~\cite{Adamatzky2007,Madikas2018}, while in~\cite{Tsuda2004} through laboratory experiments it has demonstrated Boolean logic implementation. Adamatzky and de Lacy Costello~\cite{Adamatzky2004,Costello2005} introduced a numerical simulation in which, by colliding local excitation or wave-fragments on a chemical medium, one can construct a fully functional set of logic gates. In a combination of these two collision-based computing methods~\cite{Adamatzky2012c}, Adamatzky adapted the \textit{Physarum}'s behaviour and created an experimental two-input two-output logic gates prototype~\cite{b41}.  

The aforementioned experiments require specific laboratory equipment and, usually, specialized scientific knowledge. Furthermore, given the long time intervals required for the slime mould to evolve and provide experimental result (these experiments may last from hours up to a couple of days or even more~\cite{adamatzky2012}), there is a need of reproducing its behaviour precisely and ensuring reasonable simulation time. A commonly known substitute to address these challenges is conventional Cellular Automata (CA) computational approaches~\cite{tsompanas2016}, and CA based on General Purpose Graphical Processing Units (GPGPUs) models~\cite{dourvas2016b} that reproduce plasmodium's behaviour, while achieving accurate results in significantly less time. In this way, for the purpose of further accelerating these computations, while reproducing accurately plasmodium's behaviour, attempts have been made on Field Programmable Gate Array (FPGA) implementations~\cite{Madikas2018,tsompanas2012,tsompanas2016physarum,dourvas2015} and Graphical Processing Unit (GPU) implementations~\cite{dourvas2016,8758416} that utilize hybrid CA models and enrich the expected parallelism likewise. 

\subsection{Cellular Automata computational principles revisited}

CA principles~\cite{von66,Adamatzky1995,sirakoulis2015robots} enhanced with stochastic and mainly learning abilities~\cite{narendra1974,Esnaashari2014} following the original ideas of von Neumann~\cite{von66}, can potentially introduce a hybrid, almost unconventional computational tool. It will be appropriate for dynamic systems modelling and consist of a plethora of simple components, which have the ability to learn through reinforcement and cooperate with their local surroundings to create desirable complex behaviours. The aforementioned learning concept which is considered for each cell of the CA environment, determines its internal state based on its action probability values while all the cells' evolution is dictated by the same functional update rule. The update rule and the actions chosen each time by adjusting cell's state determine its reinforcement state variable attributed by the contribution of its local adjacent neighbours. This hybrid CA environment is non-static also owing to the fact that is continuously changing as the action probability vectors of its coupling cells vary at each time step. 

The proposed computational model inspired from both CA and the relaxization of their original definition, spiced up with stochastic and learning principles, attempts to simulate the complex behaviour of \textit{P. polycephalum} to the best possible extent. The intrinsic features and computational capabilities of the proposed bio-inspired model are widely customized and verified, accordingly, through experimental data on Physarum's ability to model Logic Gates as it attempts to find minimal paths in well-formed medium with nutrients~\cite{b41}. The proposed multilevel tool is not only capable of simulating simple behaviours of the plasmodium during foraging, but also has the ability to mimic ballistic characteristics, i.e. the straightforward propagation of the plasmodium when no sources of attractants are placed on the medium, demonstrated during biological experiments~\cite{b41}. Moreover, its properly controlled learning prospect can reveal a truly stochastic behaviour, even illustrating plasmodium's weaknesses, i.e. not sufficient computations, appeared during experiments.

In the following, it will be attempted to demystify the basic characteristics of the proposed model and present them in a comprehensive and reproducible way, the main functional components of this bio-inspired model in accordance with the aforementioned basic principles of its ancestors. Further efforts will focus on the matching of the simulation results of the bio-inspired model with the experimental results of \textit{P. polycephalum} to reproduce logic gates functionality. Finally, discussion on the perspectives and enhanced abilities of the proposed model towards its application to different computing fields will conclude this paper.

\section{Physarum Modelling}

A new computational model was developed, in order to reproduce the \textit{P. polycephalum} behaviour, as the induced substrate, which constitutes the under study environment of the organism, is divided into a grid of two-dimensional (2D) identical square cells, wherein each cell corresponds to a computational unit. Each cell interacts locally via Moore’s neighbourhood, where the cell's neighbourhood is composed of the central cell $(i,j)$ and the eight adjacent cells that surround it. The state of a cell $(i,j)$ at time $t$, is defined as follows:  

\begin{equation}
Cell_{i, j}^{t}=\left\{{AS}, {Mass}_{i, j}^{t}, {PV}_{i, j}^{t}, {SD}_{i, j}^{t}, {Dir}_{i, j}^{t} {RF}_{i, j}^{t}\right\}
\end{equation} 

\noindent where \emph{AS} is a variable that defines whether that cell \textit{(i,j)} is accessible or not. Thus, this term  determines the topology of the maze. The \emph{Mass} variable represents the concentration of organism’s protoplasmic mass in the cell \textit{(i,j)}. The \emph{PV} variable is an action probability $n$-length variable, where $n$ is the size of the considered neighbourhood, that determines the probability of cell \textit{(i,j)} performing a specific action at time $t$. In particular, the cell's action corresponds to a mass transfer to a single neighbour cell and, hence, the $PV$ is an $8$-size array bearing in mind the selected Moore neighbourhood, in which each element corresponds to the probability that the cell \textit{(i,j)} will receive a portion of its neighbourhood mass. The \emph{SD} term defines the smell diffusion of the food sources and represents the smell intensity in each cell \textit{(i,j)}. Parameter \emph{Dir} indicates the direction from which neighbour mass transfer was performed at time $t$, and the \emph{RF} is a Reinforcement Flag that determines whether the action performed by the corresponding cell must be “rewarded", “penalized" or nothing should happen concerning the aforementioned action.

\begin{figure}[!hb]
  \centering
  \includegraphics[width=0.6\textwidth]{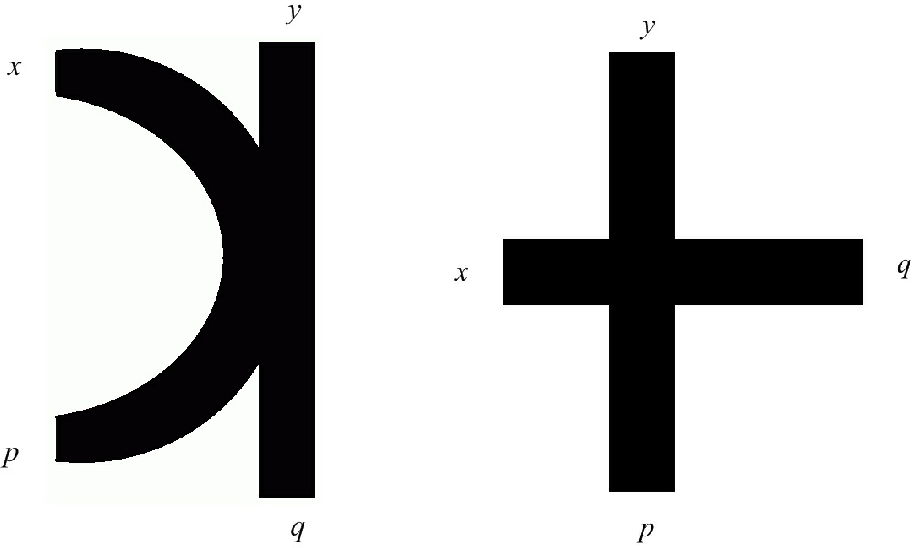}
  \caption{Geometrical structure of Physarum gates (a) $P1$ and (b) $P2$, respectively; $x$ and $y$ are inputs, while $p$ and $q$ are outputs (adopted from~\cite{b41}).}
  \label{gates}
\end{figure}

The basic parameters of the proposed model, namely the size of the two-dimensional square grid, the smell diffusion of the food, the mass portion that each cell gets from its neighbour-cells, a minimum mass threshold of the organism that is considered detectable, a minimum amount of smell threshold that is also considered detectable and a reinforcement value that determine how much a cell's energy will be strengthened or weakened should be appropriately initialized. The Probability Value ($PV$) is initialized to a uniform distribution which means practically that if the cell does not receive any kind of reinforcement, it has a \( \frac{1}{8} \) chance of receiving mass from each neighbour accordingly. Following the proposed problem definition, the organism's mass is placed at the initial input points of the logic gates $x$ and $y$, according to the gate used and the simulation carried out as depicted in Fig.~\ref{gates}. 

In the simplest scenario, the organism is allowed to diffuse freely without following any source of food and without any kind of reinforcement from its environment. This technically results in a stochastic diffusion which over time tends to look like a circle, consistent with the living organism's diffusion in the free medium. At the same time, the food smell is uniformly distributed in space according to Eq.~\ref{eq:food} and~\ref{eq:food2}. 

\begin{equation}
S D_{i, j}^{t+1}=\frac{1}{36}\sum_{i,j}\left[\begin{array}{ccc}
{1} & {4} & {1} \\
{4} & {16} & {4} \\
{1} & {4} & {1}
\end{array}\right] \cdot S D^{t}_{i, j} 
\label{eq:food}
\end{equation}

\begin{equation}
S D^{t}_{i, j}=\left[\begin{array}{ccc}
{S D_{i-1, j-1}} & {S D_{i-1, j}} & {S D_{i-1, j+1}} \\
{S D_{i, j-1}} & {S D_{i, j}} & {S D_{i, j+1}} \\
{S D_{i+1, j-1}} & {S D_{i+1, j}} & {S D_{i+1, j+1}} 
\end{array}\right]
\label{eq:food2}
\end{equation}

\noindent where $SD_{i,j}^{t+1}$ represents the Smell Diffusion on the \emph{Cell(i,j)} through its local environment (Moore's Neighbourhood). Briefly, the $SD_{i,j}^{t+1}$ updates its state after the convolution of the local surroundings $SD$ states with the kernel mentioned above.

As the organism and the smell from food sources diffuse on the two-dimensional grid, the minimum detectable mass meets the minimum detectable smell. This is the condition that activates the reinforcement rule. This rule adjusts the interactions between the cells, in order to assemble the movement of the organism towards the food. In this case when a cell $A$ transfers its mass to another adjacent cell $B$, in which the smell intensity is higher, then mass transfer from $A$ to $B$ is rewarded by the \textit{Reinforcement flag}; on the contrary, if the smell intensity is lower then it is penalized. On the next iteration, the algorithm checks the grid for reward or penalty and updates their action probability variable according to Eqs.~\ref{eq:53} and~\ref{eq:54}, respectively. 

\begin{equation}
\begin{array}{l}
{P V_{i, j, k}^{t+1}=P V_{i, j, k}^{t}+reward \cdot (1-P V_{i, j, k}^{t})} \\ \\
{P V_{i, j, k}^{t+1}=(1-reward) \cdot P V_{i, j, k}^{t}}
\end{array}  k=1,2,\ldots,n
\label{eq:53}
\end{equation}

\begin{equation}
\begin{array}{l}
{P V_{i, j, k}^{t+1}=(1-penalty) \cdot P V_{i, j, k}^{t}} \\ \\
{P V_{i, j, k}^{t+1}=\frac{penalty}{1-n}+(1-penalty) \cdot P V_{i, j, k}^{t}}
\end{array} k=1,2,\ldots,n
\label{eq:54}
\end{equation}

\noindent where \emph{reward} and \emph{penalty} take values in the range of $(0,1)$. The term \emph{k} describes the cell’s various actions and in our case represents the neighbour to which a cell transfers a percentage of its mass (as mentioned before, considering Moore’s neighbourhood here, $n=8$). 

In the case of rewarding (Eq.~\ref{eq:53}), the action's probability increases by a reward factor, while the others decrease collectively by the same factor, in order to keep the total probability of receiving any action to ‘1’. During penalty, i.e. application of Eq.~\ref{eq:54}, the exactly opposite process occurs, with the action's probability decreased by a equivalent penalty factor, while the others increase. Thus, in reward case, the \emph{Cell} is more likely to repeat the desirable action that has just happened. 

Through this repetitive learning procedure by the reinforcement from the medium, the model learns to move towards food. The intensity of this phenomenon depends heavily on the applied reinforcement factors. It is worth noting that through repeated positive or negative adjustments for the same “good" or “bad" corresponding actions, this bio-inspired model tends to follow a specific behaviour; more specifically, the probability of an action in $PV$ converges to ‘1’ and all other to ‘0’. This state is irreversible by rewards and in this model should be avoided. To alleviate such an unwanted situation there is a threshold, in analogy to the biological stochasticity of the corresponding procedures, that prevents a high unlikely action to reach a probability higher than 0.75, while keeping the rest probabilities at a minimum threshold of 0.25. This will allow the overall system to maintain its stochastic nature even under strong reinforcement. 

\begin{figure}[!h]
\centerline{\includegraphics[scale = 0.35]{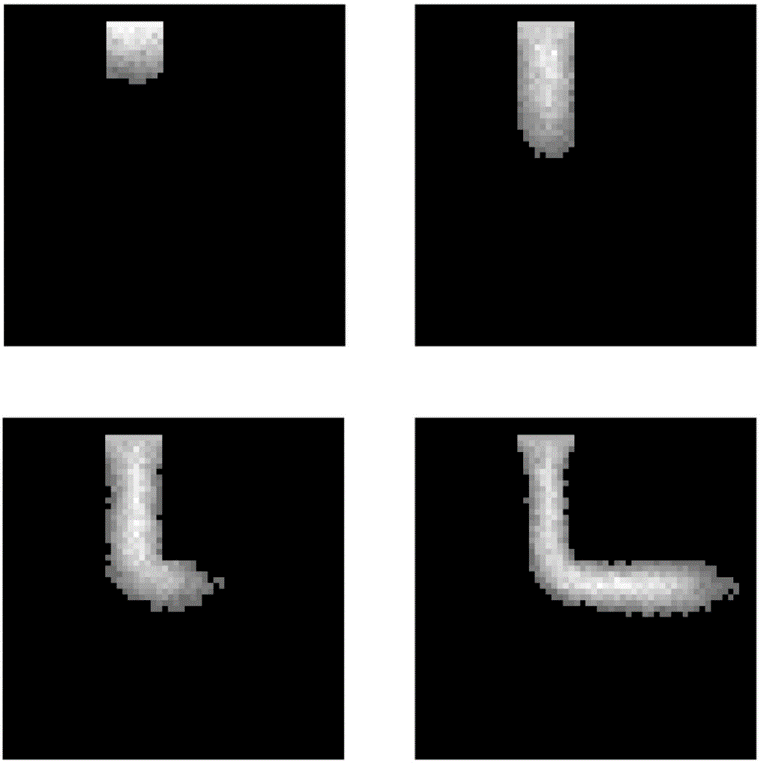}}
\caption{Diffusion of the organism under strong reinforcement for food smell. Structure of logic gate $P2$ with initial conditions (1.0) and food located east ($Q$-output) were considered.}
\label{54}
\end{figure}

An example, in accordance with the under study experiments, of such a strong reinforcement, is shown in Fig.~\ref{54} with a food source located to the east side ($q$) of $P2$ gate (Fig.~\ref{gates}) and a starting point on the north input ($y$). In this experiment, the model is configured to follow closely the smell of the food and is clearly different from diffusion in free medium. 

To simulate the ballistic behaviour of the plasmodium, a similar technique is used. We create a new scaled wave medium in which we let a wave to be diffused in the form as shown in the experiments performed in~\cite{b41} . The \emph{cells} are then rewarded from the model as they mimic the ballistic behaviour with the corresponding parameters. By releasing the reward parameter of the smell's attraction from the ballistic behaviour and generating independent reinforcements, the model is able to be controlled to an extremely precise degree. 

\section{Simulation and Experimental Results}

The grid's configuration is obtained from two-input two-output Boolean logic gates $P1$ and $P2$ depicted in Fig.~\ref{gates} and used in experiments of previous works \cite{b41}. After thorough examination and study of the tentative analogies between the experimental structures and the corresponding simulated ones, it was figured out that each gate will consist of a two-dimensional square grid of $60\times 60$ cells that will be in position to implement a transformation defined as $\langle x,y\rangle\rightarrow\langle p,q\rangle$. 

\begin{figure}[!h]
\centering
\includegraphics[width=\linewidth]{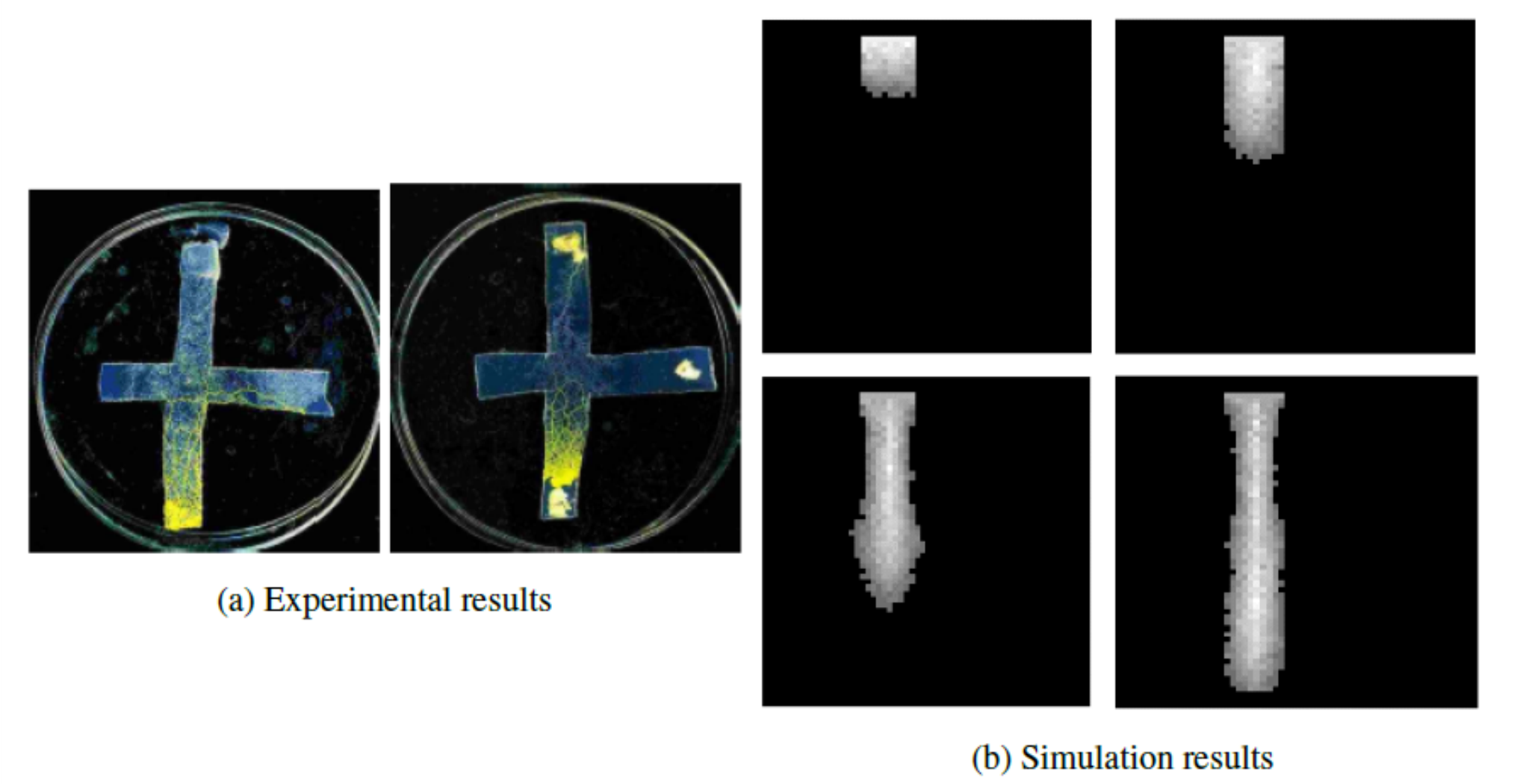}
\caption{Experimental results, adopted from \cite{b41} kindly provided from Prof. Adamatzky, and simulation results for $\langle0,1\rangle \rightarrow \langle1,0\rangle$ transformation of the proposed model on the gate $P2$.} 
\label{55}
\end{figure}

\begin{figure}[!h]
\centering
\includegraphics[width=0.8\linewidth]{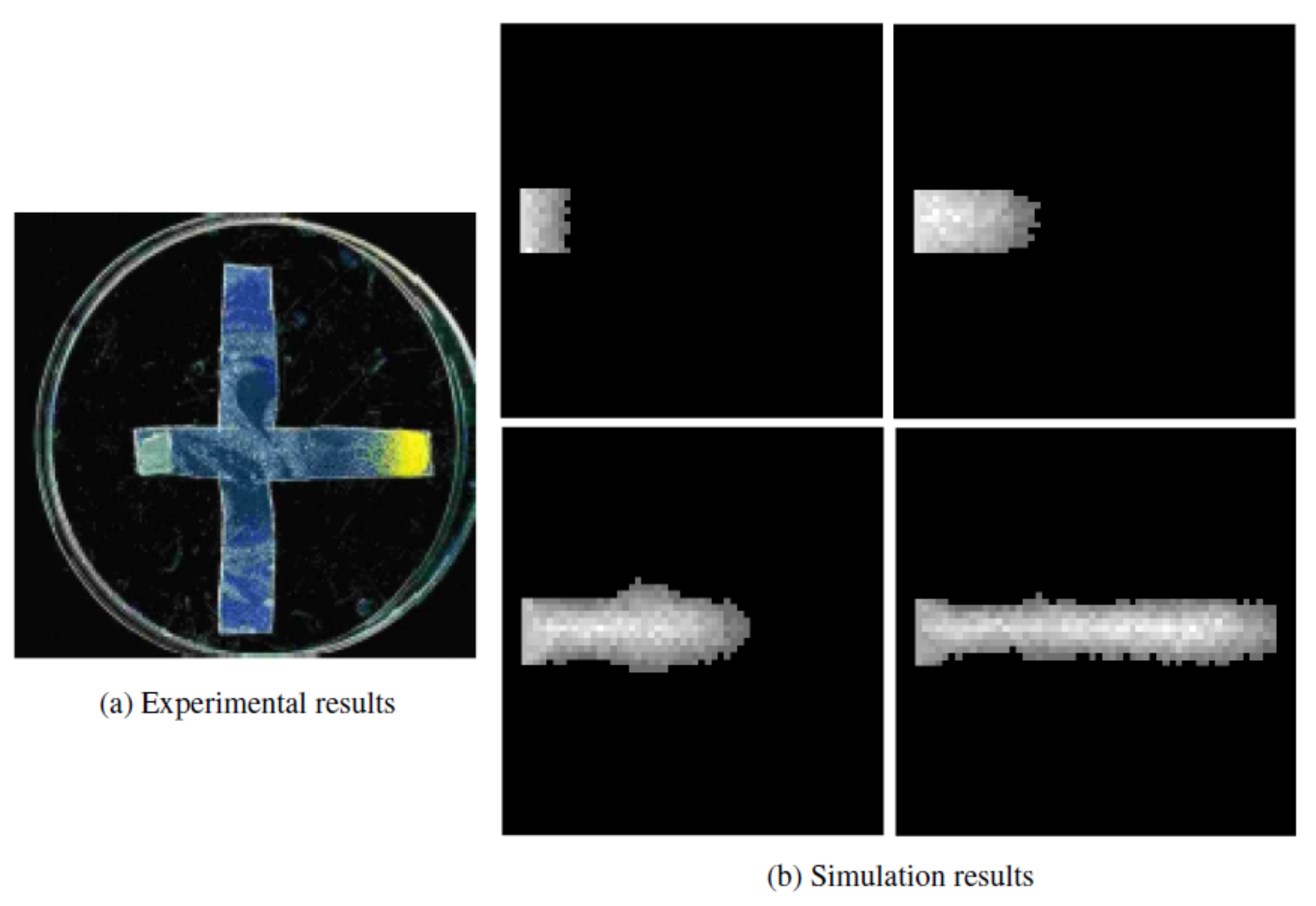}
\caption{Experimental results, adopted from \cite{b41} kindly provided from Prof. Adamatzky, and simulation results for $\langle1,0\rangle \rightarrow \langle0,1\rangle$ transformation of the proposed model on the gate $P2$.}
\label{56}
\end{figure}

\begin{figure}[!h]
\centering
\includegraphics[width=\linewidth]{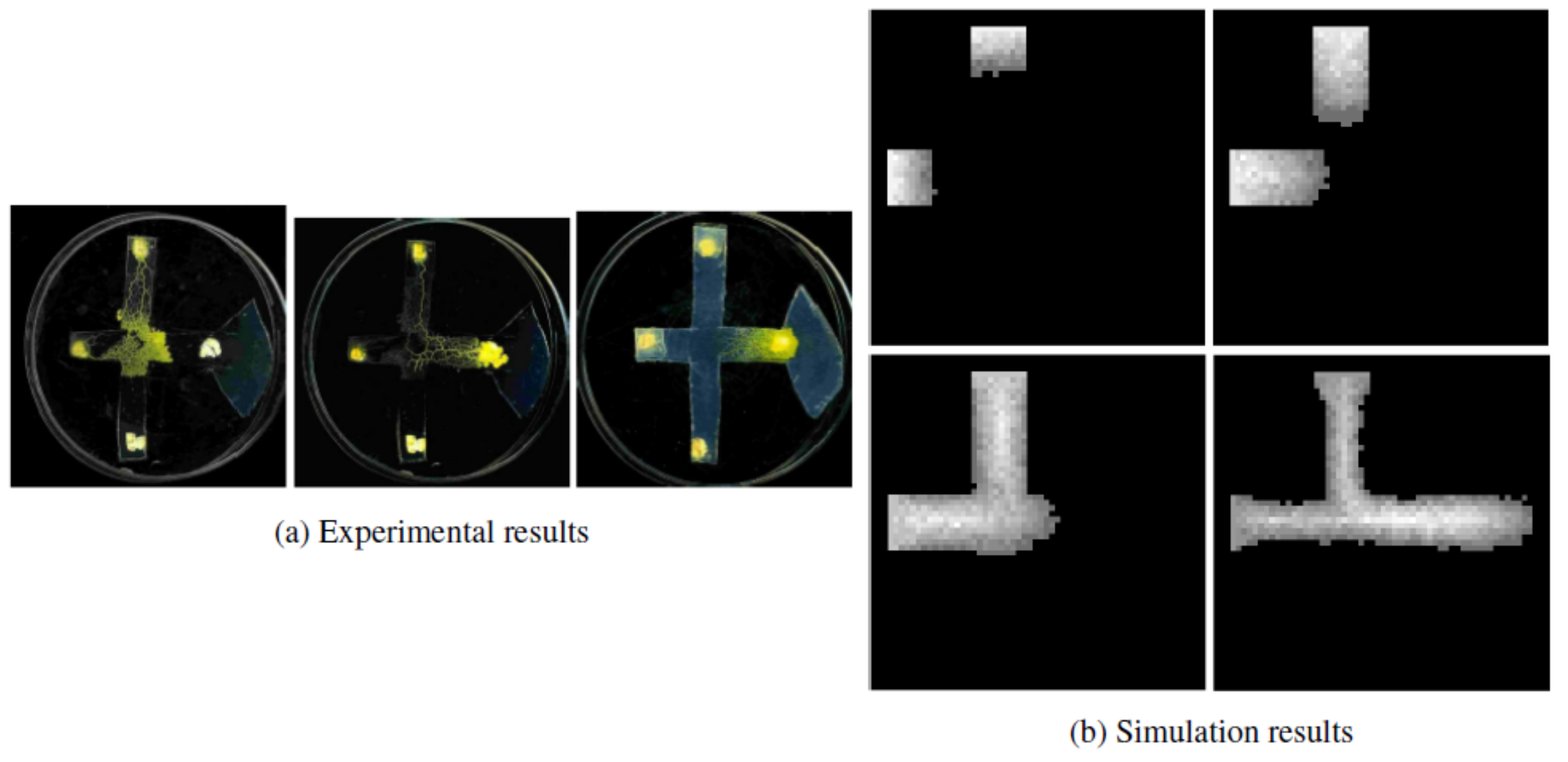}
\caption{Experimental results, adopted from \cite{b41} kindly provided from Prof. Adamatzky, and simulation results for $(\langle1,1\rangle \rightarrow \langle0,1\rangle$ transformation of the proposed model on the gate $P2$.} 
\label{57}
\end{figure}

\begin{figure}[!h]
\centering
\includegraphics[width=\linewidth]{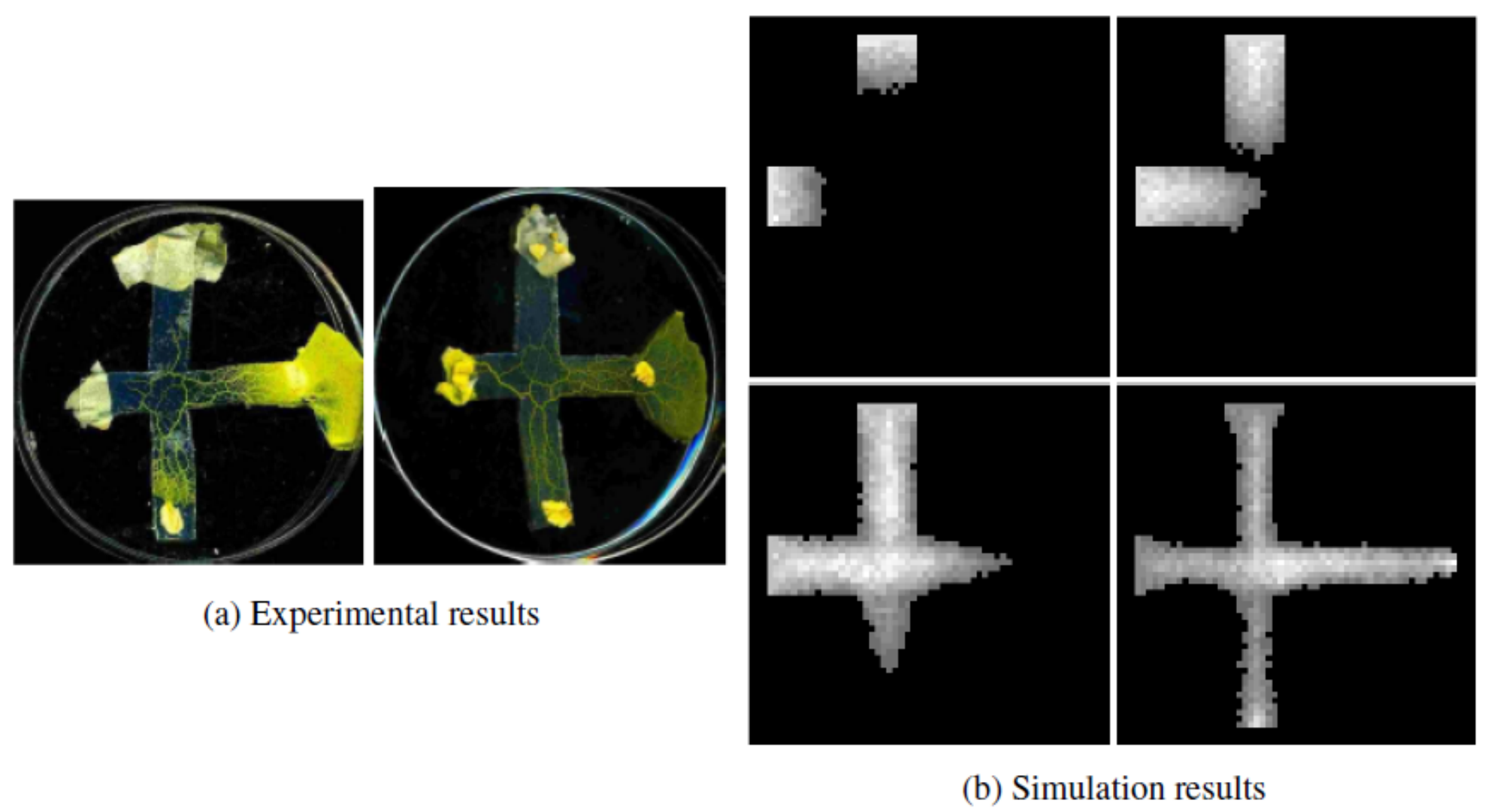}
\caption{Experimental results, adopted from \cite{b41} kindly provided from Prof. Adamatzky, and simulation results for $\langle1,1\rangle \rightarrow \langle0,1\rangle$ transformation of the proposed model on the gate $P2$ with plasmodium's presence on output $p$.}
\label{58}
\end{figure}

Figures~\ref{55} - \ref{58} exhibit the behaviour of the proposed model compared to previous experiments~\cite{b41}, for various input combinations on $P2$ gate. More specifically,  Figs.~\ref{55} and~\ref{56} illustrate the implementation of the $\langle0,1\rangle\rightarrow\langle1,0\rangle$ transformation and the  $\langle1,0\rangle\rightarrow\langle0,1\rangle$ transformation, respectively.

Figures~\ref{57} and~\ref{58} illustrate the implementation of the  $\langle1,1\rangle\rightarrow\langle0,1\rangle$ transformation. Figure~\ref{58} is of particular interest as the method of releasing the reinforcement parameters is applied. The model receives a strong enhancement in wave behaviour and a weaker feed-in magnification, located south (output $p$). By properly adjusting the two parameters, the proposed bio-inspired model manages to emulate, up to a very accurate extent, the particular behaviour of \textit{P. polycephalum}. Moreover, with similar configuration, it is possible to model all the experimental observations presented in \cite{b41}.

\begin{figure} [!h]
\centering
\includegraphics[width=0.7\linewidth]{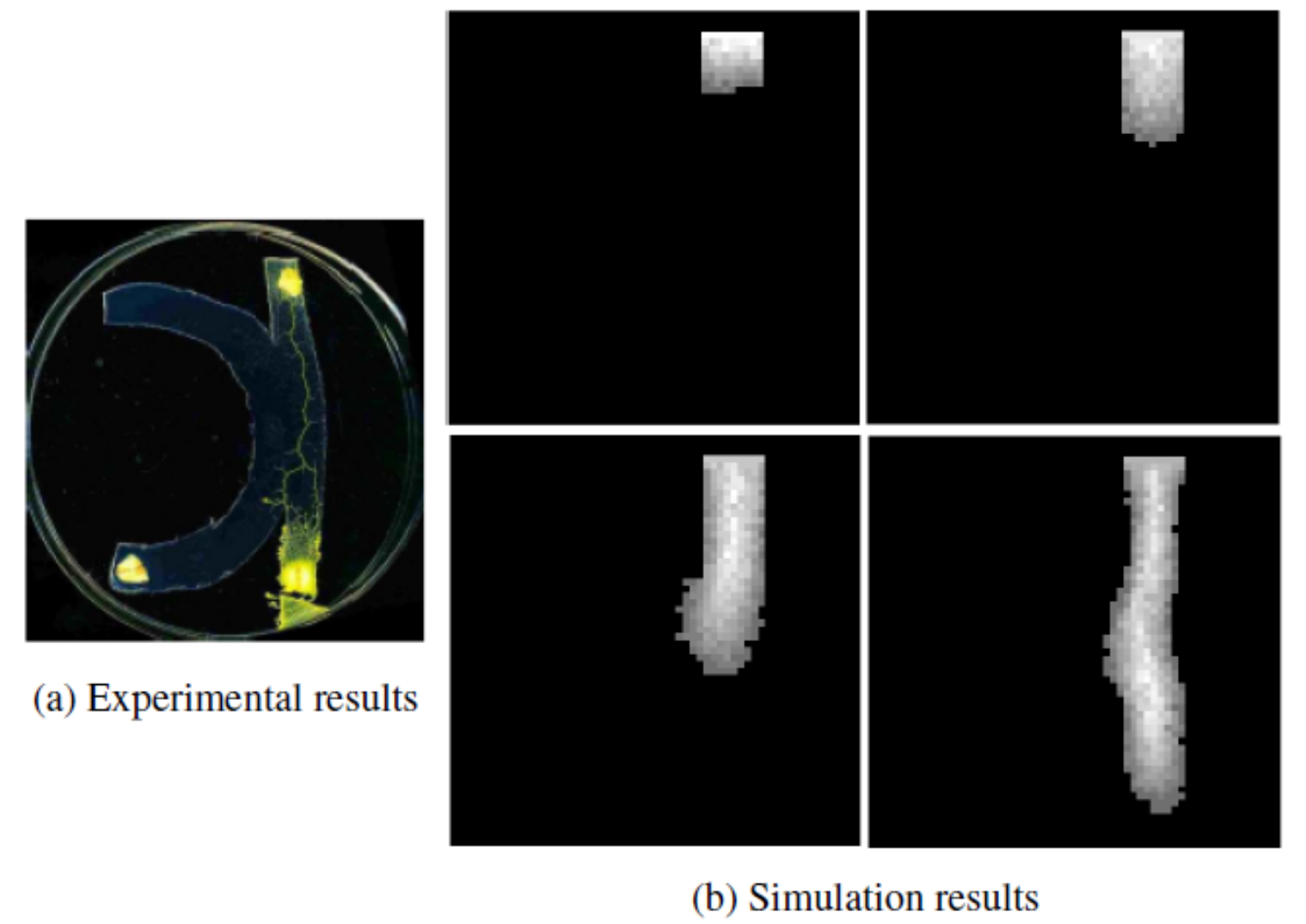}
\caption{Experimental results, adopted from \cite{b41} kindly provided from Prof. Adamatzky, and simulation results for $\langle0,1\rangle \rightarrow \langle0,1\rangle$ transformation of the proposed model on the gate $P1$.}
\label{59a}
\end{figure}

\begin{figure}[!h]
\centering
\includegraphics[width=0.7\linewidth]{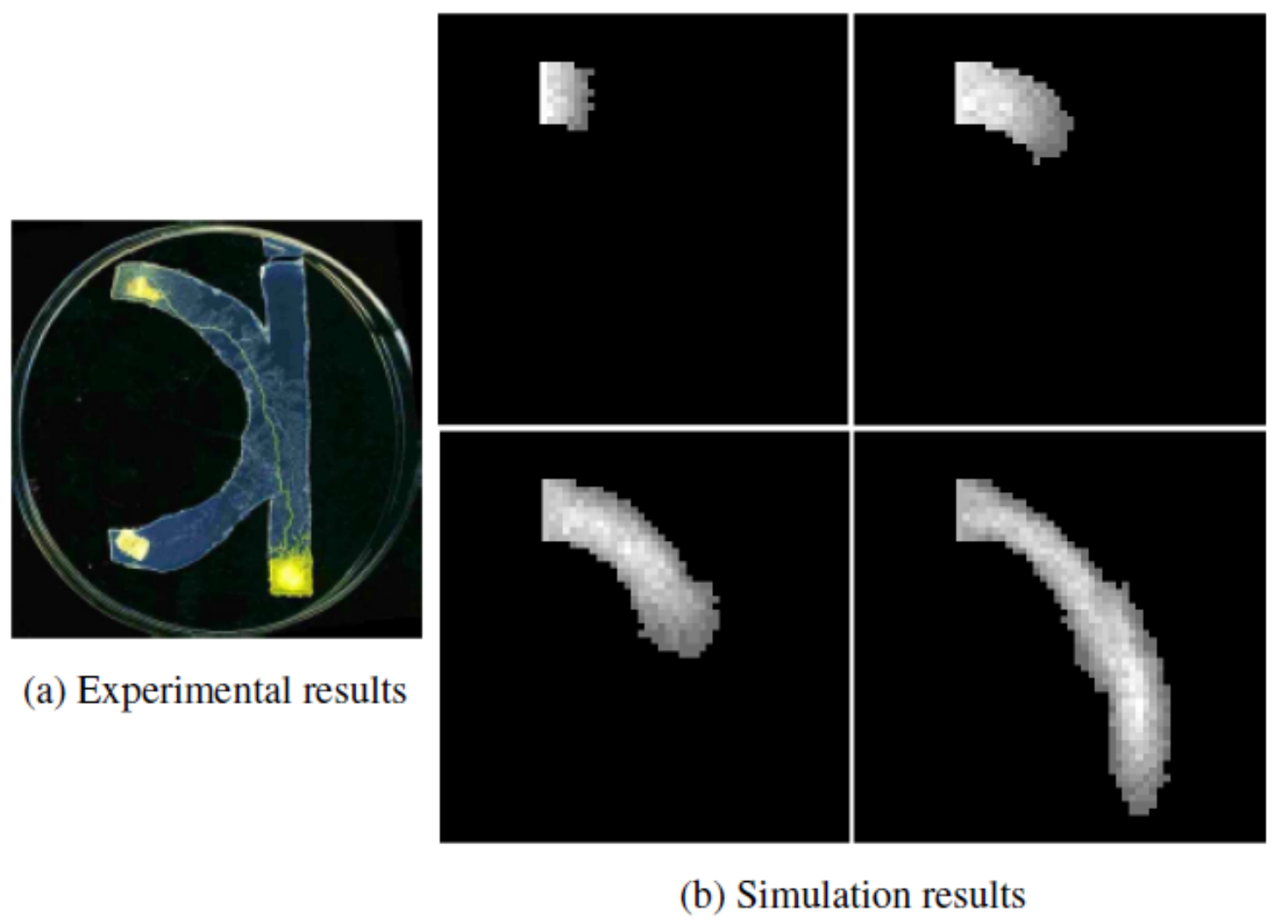}
\caption{Experimental results, adopted from \cite{b41} kindly provided from Prof. Adamatzky, and simulation results for $\langle1,0\rangle \rightarrow \langle0,1\rangle$ transformation of the proposed model on the gate $P1$.}
\label{59b}
\end{figure}

\begin{figure} [!h]
\centering
\includegraphics[width=0.9\linewidth]{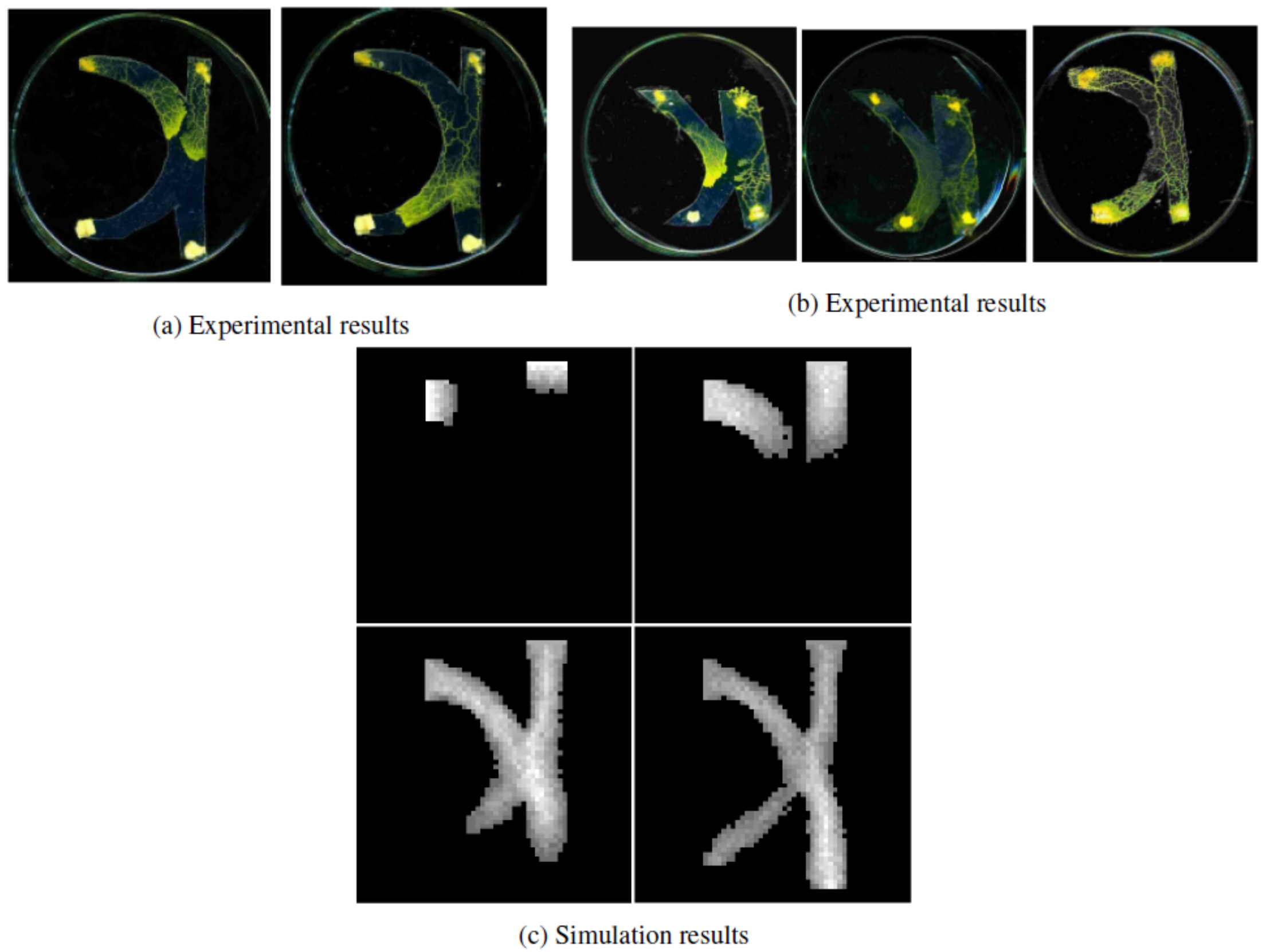}
\caption{Experimental results, adopted from \cite{b41} kindly provided from Prof. Adamatzky, and simulation results for $\langle1,1\rangle \rightarrow \langle1,1\rangle$ transformation of the proposed model on the gate $P1$.}
\label{510}
\end{figure}

Figures~\ref{59a} - \ref{510} exhibit the behaviour of the proposed model for various input combinations on $P1$ gate. In specific, Fig.~\ref{59a} and~\ref{59b} illustrate the $\langle0,1\rangle\rightarrow\langle0,1\rangle$ transformation and the $\langle1,0\rangle\rightarrow\langle0,1\rangle$ transformation, respectively. These are the cases where one input of the gate is activated, which has the same output (activation of output $q$), as proved in biological experiments and successfully simulated by the model. Lastly, the case of both inputs activated was also studied and the results are depicted in Fig.~\ref{510}. 

It is evident that the introduced bio-inspired model is properly sufficient to describe the particular laboratory details while it produces accurate simulation results that closely follows in different time steps and are in appropriate agreement with the experimental results. The frequency of occurrence of these results also matched the experimental data, but at this point it should be noted that the model parameters were determined in such a way as to derive this agreement. As demonstrated by the case of the transformation $\langle1,1\rangle\rightarrow\langle0,1\rangle$ on the $P2$ gate (Fig.~\ref{57} -- Fig.~\ref{58}) with suitable configuration of the model, it is feasible to exhibit various behaviours, even under the same initial conditions. In addition to fully deterministic behaviour, originated by the CA principles, the presented model is capable of being fully stochastic, as well as, it can demonstrate any probability distribution, including that observed by Adamatzky in \cite{b41} in the corresponding laboratory experiments.

\section{Conclusions}
In this paper, an almost unconventional bio-inspired model drawing inspiration from the inherent parallelism of Cellular Automata computational model and their ability to provide adequate solutions to model complex physical phenomena and processes, is further enriched from the stochasticity of Learning Automata and the corresponding learning abilities. 

The proposed model was selected to be stressed in an ambitious and of high complexity modeling project, namely modeling of living organisms and their properties. The plasmodium of \textit{P. polycephalum} owing to the fact that although is formulated as a single-cell, it serves as a prosperous bio-computational example for a wide range of real life applications. As the presented simulation results indicate, the model developed is capable of modeling very accurately the computing behaviour of \textit{P. polycaphalum} and its ballistic nature, as well as, its unique development in foraging for a number of selected cases. More specifically, the testbed selected was a series of biological experiments, as provided in \cite{b41}, so as the \textit{P. polycephalum} to perform basic logical operations on a geometrically constrained substrate of two Boolean logic gates, where the proposed model performed adequately mimicking in an efficient way the biological organism behaviour. Up-to-date models have attempted to imitate this behaviour through either less or more complicated CA rules \cite{tsompanas2016cellular,tsompanas2016physarum,tsompanas2014} or other methods \cite{tero2008flow,Tero2010,gunji2011adaptive,jones2014computation,zhang2015biologically,wu2015new,zhang2013biologically,liu2019simulating,gao2018does}; however, the flexibility and the precision offered by the introduced model, through almost simpler rules, makes it also an excellent tool for modeling such organisms.

As have been proven, the proposed model can be applied to a plenty of high complexity problems by properly adjusting the corresponding rules and the initial conditions to further boost the model's performance when trying to meet with the experimental data of the under study problem, in a qualitative and quantitative manner. Taking into account the results presented in this paper, it can be foreseen that the proposed bio-inspired model can serve as an efficient tool in further studies to model the behaviour of other, even more complex, living organisms as well as in many similar problems.

\bibliographystyle{unsrt}

\end{document}